\documentstyle[multicol,prc,aps,eqsecnum,epsfig]{revtex}
\begin{document}
\draft

\title{
	$^{92}$Mo($\alpha$,$\alpha$)$^{92}$Mo scattering,
	the $^{92}$Mo--$\alpha$ optical potential, and
	the $^{96}$Ru($\gamma$,$\alpha$)$^{92}$Mo reaction rate
	at astrophysically relevant energies
}

\author{
Zs.~F\"ul\"op,
Gy.\ Gy\"urky,
Z.~M\'at\'e,
E.~Somorjai,
L.~Zolnai
}
\address{
	Institute of Nuclear Research of the Hungarian Academy 
	of Sciences, PO Box 51, H-4001 Debrecen, Hungary
}

\author{
D.~Galaviz\footnote{corresponding author, email: Redondo@ikp.tu-darmstadt.de},
M.~Babilon,
P.~Mohr,
A.~Zilges
}
\address{
	Institut f\"ur Kernphysik, Technische Universit\"at Darmstadt,
	Schlossgartenstrasse 9, D--64289 Darmstadt, Germany
}

\author{
T.~Rauscher
}

\address{
	Departement f\"ur Physik und Astronomie, Universit\"at Basel, 
	Klingelbergstrasse 82, CH-4056 Basel, Switzerland; and
	  Department of Astronomy and Astrophysics, University of
        California, Santa Cruz, CA 95064, USA
}

\author{
H.~Oberhummer
}
\address{
	Atominstitut of the Austrian Universities, 
	Vienna University of Technology,
	Wiedner Hauptstr.~8-10, A--1040 Vienna, Austria
}

\author{
G.~Staudt
}
\address{
        Physikalisches Institut, Universit\"at T\"ubingen,
        Auf der Morgenstelle 14, D--72076 T\"ubingen, Germany
}

\date{\today}

\maketitle

\begin{abstract}
The elastic scattering cross section of
$^{92}$Mo($\alpha$,$\alpha$)$^{92}$Mo has been measured at energies of
$E_{\rm{c.m.}} \approx$ 13, 16, and 19\,MeV in a wide angular
range. The real and imaginary parts of the optical potential for the
system $^{92}$Mo - $\alpha$ have been derived at energies around and
below the Coulomb barrier. The result fits into the systematic
behavior of $\alpha$-nucleus folding potentials. 
The astrophysically relevant
$^{96}$Ru($\gamma$,$\alpha$)$^{92}$Mo reaction rates
at $T_9=2.0$ and $T_9=3.0$ could be determined to an accuracy of about
16\,\% and are compared to previously published theoretical rates.
\end{abstract}


% 24.10.Ht: Optical and diffraction models
% 25.55.-e Sup 3 H-, sup 3 He-, and sup 4 He-induced reactions
% 25.55.Ci Elastic and inelastic scattering
% 26.30.+k Nucleosynthesis in novae, supernovae and other explosive
%          environments

\begin{multicols}{2}
\narrowtext

\section{Introduction}
\label{sec:intro}
The nucleosynthesis of nuclei above the iron peak ($A \approx 60$)
proceeds mainly by neutron capture in the so-called $s$- and
$r$-processes. In principle, neutron-deficient nuclei in this mass region (the
so-called $p$ nuclei, see Ref.~\cite{Lam92} for a complete list) can
be produced from more neutron-rich seed nuclei either by the removal of
neutrons or by the addition of protons. However, due to the Coulomb barrier
proton capture is strongly suppressed.
There is general agreement that heavy neutron-deficient nuclei with
masses above $A \approx 100$ have been synthesized by photodisintegration
of previously produced nuclides
at sufficiently high temperatures of $(2-3) \times 10^9$\,K
($T_9 = 2-3$, with $T_9$ being the temperature in
billion degrees). This so-called $\gamma$-process or p-process is discussed in
detail in 
\cite{Lam92,Arn99,Lan99,Wal97,Ito61,Mac70,Woo78,Ray90,Ray95,Pra90,How91,Cos2000}.
Several astrophysical sites 
for the $\gamma$-process have been proposed, 
whereby the oxygen- and neon-rich layers of type II supernovae
seem to be good candidates \cite{Woo78,Pra90}, but also exploding
carbon-oxygen white dwarfs have been suggested \cite{How91}. However, no
definite conclusions have been reached yet.

Nucleosynthesis calculations for the $\gamma$-process require a huge
number of reaction rates. Up to about 1000 nuclei and 10000 reaction
rates have been included in previous reaction networks \cite{Ray90}. 
Recently, the complete network of the first self-consistent study of the 
$\gamma$-process including all relevant nuclei up to Bi amounted to
about 3000 nuclei and all their respective reactions \cite{heg00,rau_apj01}.
Unfortunately, almost none of these reaction rates have
been measured and the astrophysical calculations have to rely completely
on statistical model calculations (e.g.~Refs.~\cite{Rau2000,Gor2000,Rau2001}).
Of special importance are ($\gamma$,$\alpha$)/($\gamma$,n) branchings
which determine abundance ratios of certain nuclides which, in turn, can
in some cases be compared to abundances found in meteoritic 
inclusions \cite{wh90,rau95,som98}. 
It has been stated that the uncertainties for ($\gamma$,$\alpha$)
reaction rates are huge \cite{Gor2000,wh90,rau98}. 
The determination of the
$\alpha$-nucleus potential at astrophysically relevant energies helps
to reduce the uncertainties of the calculations significantly
\cite{Mohr97,gle00}.
($\gamma$,n) reaction rates have been measured in a recent work using a
quasi-thermal photon spectrum, and rough agreement between theoretical
predictions and the measured rates was found \cite{Mohr00,Vogt01}.

The overall agreement between the calculated and the observed
abundance patterns of the $p$ nuclei is relatively good. However,
the mass region $70 \le A \le 100$ is generally underproduced
in the nucleosynthesis calculations
\cite{Woo78,Cos2000,heg00}. The production among others
depends on the neutron-producing
$^{22}$Ne($\alpha$,n)$^{25}$Mg reaction rate (which may enhance the
$s$-process seed nuclei for the $\gamma$-process \cite{Cos2000}) and
on the photodisintegration rates
in the $\gamma$-process but it remains unclear whether the
underproduction can be cured by a change in those rates \cite{woo2001}.
Other explanations for this discrepancy include additional production
mechanisms like neutrino-induced nucleosynthesis \cite{Hoff96} and
additional production sites like the rapid proton capture ($rp$)
process on accreting neutron stars (e.g.\ \cite{Schatz2000,Wal81}).
However, it is still an open question whether $rp$-material can be
ejected into the interstellar medium in sufficient quantities from these
X-ray bursters \cite{Schatz2001}.

\begin{figure}
\centerline{\epsfig{file=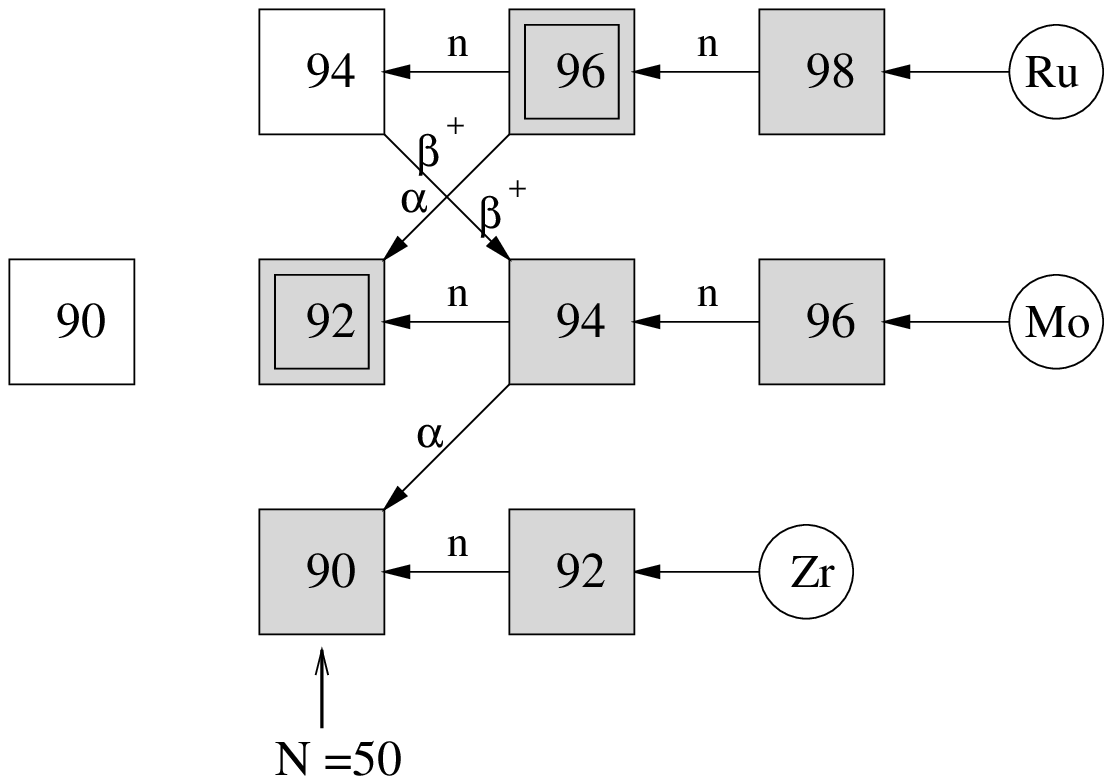,angle=0,width=7cm}}
\caption{
Nucleosynthesis path for $^{92}$Mo and $^{94}$Mo in the astrophysical
$\gamma$-process. Stable nuclei are gray shaded. The nuclei $^{92}$Mo
and $^{96}$Ru are marked; in this paper we determine an improved
reaction rate for the $^{96}$Ru($\gamma$,$\alpha$)$^{92}$Mo
reaction. Note that the
($\gamma$,n) reactions stop at the magic neutron number $N = 50$. The circles 
at the right end of the diagram mark several other stable nuclei.
}
\label{fig:gammaproc}
\end{figure}

The motivation of the experimental determination of the 
$\alpha$-nucleus potential for $^{92}$Mo is twofold. The
determination of the $\alpha$-nucleus potential at energies below the
Coulomb barrier is limited in general because ($i$) the experimental data show
only small deviations from the Rutherford cross section, and ($ii$)
the optical potentials have ambiguities. An experiment on $^{92}$Mo
allows to extend the systematic study of $\alpha$-nucleus potentials
\cite{Mohr97,gle00,Atz96,Mohr2000a} to lower energies. The second motivation 
refers directly to the production of the $p$ nuclei $^{92}$Mo and
$^{94}$Mo. A possible reaction path leading to the production of
$^{92}$Mo and $^{94}$Mo is shown in
Fig.~\ref{fig:gammaproc}. Photodisintegration reactions of the nucleus
$^{96}$Ru can lead to the production of ($i$) $^{92}$Mo by
$^{96}$Ru($\gamma$,$\alpha$)$^{92}$Mo and ($ii$) $^{94}$Mo by
$^{96}$Ru($\gamma$,n)$^{95}$Ru($\gamma$,n)$^{94}$Ru(2\,$\times$
$\beta^+$)$^{94}$Mo. If this reaction path were the only production
mechanism for $^{92}$Mo and $^{94}$Mo, the abundance ratio between
$^{92}$Mo and $^{94}$Mo would be directly related to the ratio of
($\gamma$,$\alpha$) and ($\gamma$,n) reaction rates of
$^{96}$Ru. In this case the ratio $^{94}$Mo/$^{92}$Mo could be a thermometer 
for the $\gamma$-process because of the temperature dependence of the 
($\gamma$,n)/($\gamma$,$\alpha$) branching ratio. However, for a quantitative 
analysis contributions of the $rp$-process to $^{92}$Mo and $^{94}$Mo and the 
weak $s$-process contribution to $^{94}$Mo have to be known.

The choice of the measured energies at about 13, 16, and 19\,MeV has
the following reasons. The astrophysically relevant energy window for
($\gamma$,$\alpha$) reactions at $T_9 \approx 2-3$ is of the order of
$E_\gamma \approx 8-10$\,MeV corresponding to $6-8$\,MeV for the
reverse reaction $^{92}$Mo($\alpha$,$\gamma$)$^{96}$Ru. Scattering
experiments at these low energies are possible; however, a reliable
determination of optical potentials is impossible because of the
dominating Coulomb interaction. The height of the Coulomb barrier is
about 15 MeV. We decided to measure at several energies above and below the
Coulomb barrier to extract the optical potential and its energy
dependence at energies as close as possible to the astrophysically
relevant energy range.

In the following paper we first present our experimental setup
(Sect.~\ref{sec:exp}). The experimental results are analyzed using
systematic folding potentials, and discrete and continuous ambiguities
are discussed in detail (Sect.~\ref{sec:OM}). The optical
potential at astrophysically relevant energies is determined by
extrapolation using the systematic behavior of $\alpha$-nucleus
potentials \cite{Atz96,Mohr2000a}, and the ($\gamma$,$\alpha$)
reaction rates are calculated (Sect.~\ref{sec:extra}). Finally, some 
conclusions are given (Sect.~\ref{sec:summ}). A preliminary
analysis of this experiment was presented already in \cite{MohrNIC}.

\section{Experimental Setup and Procedure}
\label{sec:exp}
The experiment was performed at the cyclotron laboratory at ATOMKI,
Debrecen. We used the $78.8~{\rm{cm}}$ diameter scattering chamber
which is described in detail in Ref.~\cite{Mate89}.  Here we 
discuss only those properties which are important for the present
experiment. A similar setup has been used in our previous
$^{144}$Sm($\alpha$,$\alpha$)$^{144}$Sm experiment \cite{Mohr97}.

\subsection{Targets}
\label{subsec:target}
\begin{figure}
\centerline{\epsfig{file=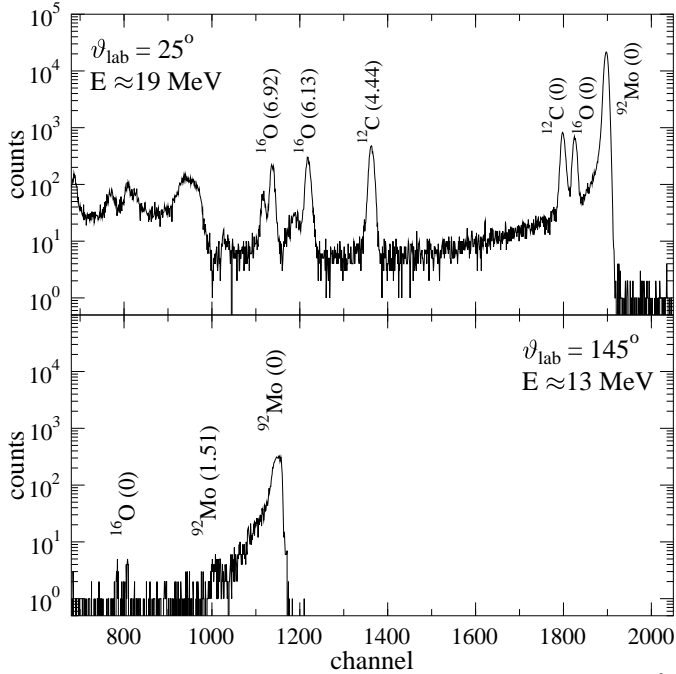,angle=0,width=9cm}}
\caption{
Typical spectra at $E \approx 19$\,MeV and $\vartheta_\alpha =
25^\circ$ (upper diagram) and $E \approx 13$\,MeV and
$\vartheta_\alpha = 145^\circ$ (lower diagram).
The peak from elastic $^{92}$Mo-$\alpha$ scattering is well
separated from $^{12}$C-$\alpha$, $^{16}$O-$\alpha$ elastic scattering (upper)
and from inelastic $^{92}$Mo - $\alpha$ scattering (lower diagram).
}
\label{fig:spectra}
\end{figure}

The $^{92}$Mo targets were produced by evaporation of 97.33\,\%
enriched $^{92}$MoO$_3$ on a thin (20\,$\mu$g/cm$^2$) carbon backing
directly before the beamtime at the target laboratory of ATOMKI. 
The target was mounted on the target holder in the center of
the scattering chamber.  The
surface of the evaporated $^{92}$MoO$_3$ turned out to be not very
flat leading to relatively broad low-energy tails in the spectra of
the elastically scattered $\alpha$ particles (see
Fig.~\ref{fig:spectra}). The target which was used during the whole
experiment had a thickness of about 200\,$\mu$g/cm$^2$. The target
stability was monitored during the whole experiment to avoid
systematic uncertainties from changes in the target foil.

\subsection{Scattering Chamber}
\label{subsec:chamber}
A remote-controlled target ladder was placed in the center of the
scattering chamber. Additionally, two apertures were mounted on the
target holder to check the beam position and the size of the beam spot
directly at the position of the target. The two apertures had a width
and height of 2x6 mm$^2$ and 6x6 mm$^2$, respectively. The apertures
were placed at the target position instead of the $^{92}$MoO$_3$ target
before and after each variation of the beam energy and beam
current. The beam was optimized until no current could be measured on
the larger aperture, and the current on the smaller aperture was
minimized (typically less than 1\,nA compared to about 300\,nA beam
current). The width of the beam spot was smaller than 2 mm during the
whole experiment, which is very important for the precise
determination of the scattering angle. Note that the relatively poor
determination of the height of the beam spot does not disturb the
claimed precision of the scattering angle (see
Sect.~\ref{subsec:angle}).

\subsection{Detectors and Data Acquisition}
\label{subsec:detector}
For the measurement of the angular distribution we used four silicon
surface-barrier detectors with an active area $A = 50~{\rm{mm^2}}$ and
thicknesses between $D = 300~{\rm{\mu m}}$ and $D = 1500~{\rm{\mu
m}}$.  The detectors were mounted on an upper and a lower turntable,
which can be moved independently. On each turntable two detectors were
mounted at an angular distance of $10^\circ$. Directly in front of the
detectors apertures were placed with the dimensions 1.25\,mm $\times$ 5.0\,mm
(lower detectors) and 1.0\,mm $\times$ 6.0\,mm (upper detectors). Together with
the distance from the center of the scattering chamber $d =
195.6~{\rm{mm}}$ (lower detectors) and $d = 196.7~{\rm{mm}}$ (upper
detectors) this results in solid angles from $\Delta \Omega = 1.63
\times 10^{-4}$ to $\Delta \Omega = 1.55 \times 10^{-4}$. The ratios
of the solid angles of the different detectors were determined
by overlap measurements with an accuracy better than 1\,\%. 

Additionally, two detectors were mounted at the wall of the scattering
chamber at a distance of $d = 351.3$\,mm and
at a fixed angle of $\vartheta = 15^\circ$ (left and right side
relative to the beam direction). These detectors were used
as monitor detectors to normalize the measured angular distribution
and to determine the precise position of the beam spot on the target.
The solid angle of both monitor detectors
is $\Delta \Omega = 8.10 \times 10^{-6}$. 

The signals from all detectors were processed using charge-sensitive
preamplifiers (PA), which were mounted directly at the scattering chamber.
The output signal was further amplified by a main amplifier (MA) and
fed into ADCs. The data were collected using the commercially
available system WinTMCA which provides an automatic deadtime control 
which was found to be reliable in a previous experiment
\cite{Mohr2000b}. For the coincidence measurements
(Sect.~\ref{subsec:angle}) additionally the bipolar signals of the MAs
were fed into Timing Single Channel Analyzers (TSCA), and the unipolar
outputs of the MAs were gated using Linear Gate Stretchers (LGS). 

The energy resolution of the detectors was tested before the
experiment using a mixed $\alpha$ source and values better than
20\,keV were measured. During the experiment the achieved energy
resolution is determined mainly by the energy spread of the primary
beam and by the thickness and flatness of the target. Depending on the
measured angle, the achieved energy resolution was between 0.5\,\% and
2\,\% corresponding to $\Delta E \approx 200~{\rm{keV}}$ at
$E_{\alpha} \approx 20~{\rm{MeV}}$. Two typical spectra are shown in
Fig.~\ref{fig:spectra}. The relevant peaks from elastic
$^{92}$Mo-$\alpha$ scattering are well separated from inelastic and
from background peaks.

\subsection{Angular Calibration}
\label{subsec:angle}
The angular calibration of the setup is of crucial importance for the
precision of a scattering experiment at energies close to the Coulomb
barrier because the Rutherford cross section depends sensitively on
the angle with $\sin^{-4}{(\vartheta/2)}$.  A small uncertainty of
$0.1^\circ$ in the determination of $\vartheta$ leads to a cross
section uncertainty of 2.0\,\% (1.0\,\%; 0.6\,\%) at an angle
$\vartheta = 20^\circ$ ($40^\circ$; $60^\circ$). The following methods
were applied to measure the precise scattering angle $\vartheta$.

The position of the beam on the target was continuously controlled by
two monitor detectors. The precise position of the beam spot was
derived from the ratio of the count rates in both monitor detectors.
Typical corrections were smaller than one millimeter,
leading to corrections in $\vartheta$ of the order of
$0.1^\circ$. However, because of the minor beam quality
at the 16\,MeV measurement, larger corrections had to be applied
to this angular distribution leading to larger uncertainties in the
determination of the optical potential.

The position of the four detectors was calibrated using the steep
kinematics of $^{1}$H($\alpha$,$\alpha$)$^{1}$H scattering at forward
angles ($10^\circ < \vartheta_{\rm{lab}} < 15^\circ$) \cite{Mohr97}. The
results of our previous experiment could be confirmed
within the uncertainties \cite{Mohr97}.

\begin{figure}
\centerline{\epsfig{file=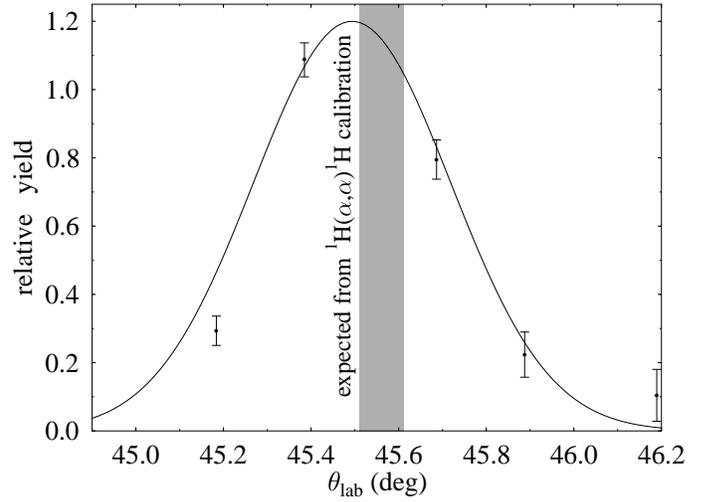,angle=0,width=9cm}}
\caption{
Relative yield of $^{12}$C recoil nuclei in coincidence with elastically
scattered $\alpha$ particles. The shaded area shows the angle and the
uncertainty which is expected from the calibration using the steep
kinematics of $^{1}$H($\alpha$,$\alpha$)$^{1}$H. The dotted line is a
Gaussian fit to the experimental data points to guide the eye.
}
\label{fig:coin}
\end{figure}

Finally, we measured a kinematic coincidence between elastically
scattered $\alpha$ particles and the corresponding $^{12}$C recoil
nuclei using a carbon backing without molybdenum as target. 
One detector was placed at $\vartheta_{\rm{lab},\alpha} =
70^\circ$ (left side relative to the beam axis), and the
signals from elastically scattered $\alpha$ particles on $^{12}$C were
selected by a TSCA. This TSCA output was used as gate for
the signals from another detector which was moved around the
corresponding $^{12}$C recoil angle $\vartheta_{\rm{lab},recoil} =
45.5^\circ$ (right side).  The maximum recoil count rate was
found almost exactly at the expected angle (see Fig.~\ref{fig:coin}).

In summary, 
the overall uncertainty of the angles $\vartheta$ in this experiment
is about $0.1^\circ$ for the measurements at 13 and 19\,MeV and about
$0.2^\circ$ for the 16\,MeV measurement.

\subsection{Experimental Procedure and Data Analysis}
\label{subsec:proc}
Three angular distributions have been measured at energies of
$E_\alpha = 19.50$, 16.42, and 13.83\,MeV. 
The beam current was between $80$\,nA and $320$\,nA. The
experiment covers the full angular range from forward angles of
$\vartheta = 20^\circ$ to backward angles of $\vartheta = 170^\circ$
in steps of about $1^\circ$ at all three energies.
The statistical uncertainties of each data
point vary from $\le 0.1\,\%$ at forward angles to about $1\,\%-2\,\%$
at backward angles.

The count rates $N(\vartheta)$ in the four detectors have
been normalized to the number of counts in the monitor detectors
$N_{\rm{Mon}}(\vartheta = 15^\circ)$:
\begin{equation}
\Bigl(\frac{d\sigma}{d\Omega}\Bigr)(\vartheta) =
	\Bigl(\frac{d\sigma}{d\Omega}\Bigr)_{\rm{Mon}}(\vartheta=15^\circ) 
	\cdot
	\frac{N(\vartheta)}{N_{\rm{Mon}}(\vartheta=15^\circ)} \cdot
	\frac{\Delta \Omega_{\rm{Mon}}}{\Delta \Omega}
\end{equation}
with $\Delta \Omega$ being the solid angles of the detectors. These
measured cross sections have been transferred to the center-of-mass
system. The cross section at the monitor position
$\vartheta_{\rm{Mon}} = 15^\circ$ is given by the Rutherford cross
section. The relative measurement eliminates the typical uncertainties
of absolute measurements which come mainly from changes in the target
and from the beam current integration. Nevertheless, the beam current
was measured by standard current integration in the Faraday cup, and
the absolute value of the cross section was consistent with the
measured relative cross sections.

\begin{figure}
\centerline{\epsfig{file=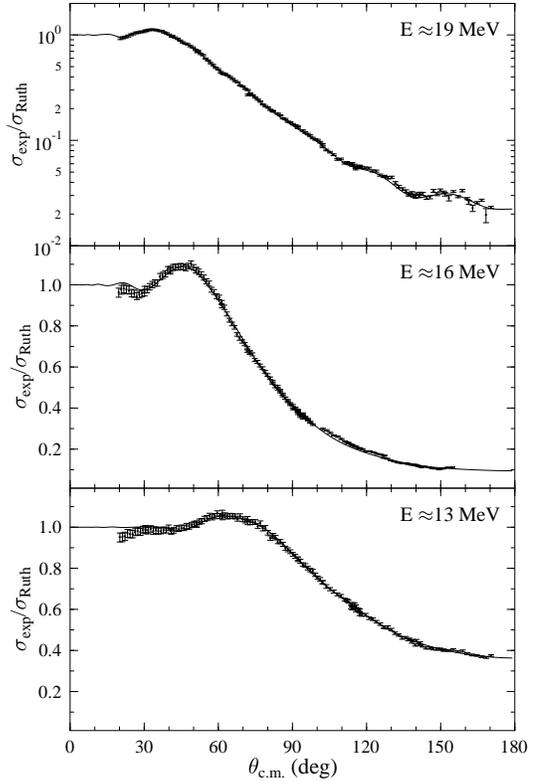,angle=0,width=7cm}}
\caption{
Experimental cross sections of $^{92}$Mo($\alpha$,$\alpha$)$^{92}$Mo
at $E_{\rm{c.m.}} \approx 13$, 16, and 19\,MeV normalized to the
Rutherford cross section. The lines are the result of optical model
calculations (see table~\ref{tab:final}).
}
\label{fig:results}
\end{figure}

The three angular distributions are shown in Fig.~\ref{fig:results}.
The lines are the result of optical model calculations  (see
Sect.~\ref{sec:OM}). The measured cross sections cover five orders of
magnitude between the highest (forward angles at $E =
13$\,MeV) and the smallest cross section (backward angles at $E =
19$\,MeV) with almost the same accuracy. Further details of the
experimental set-up and the data analysis can be found in
Ref.~\cite{Gal01}.

\section{Optical Model Analysis}
\label{sec:OM}
The theoretical analysis of the scattering data was performed in the
framework of the optical model (OM). 
The complex optical potential is given by
\begin{equation}
U(r) = V_{\rm C}(r) + V(r) + iW(r)
\end{equation}
where $V_{\rm C}(r)$ is the Coulomb potential, and $V(r)$ and $W(r)$ are the
real and the imaginary part of the nuclear potential, respectively.

\subsection{The Folding Potential}
\label{subsec:fold}
The real part of the optical potential was calculated by a
double--folding procedure:
\begin{equation}
V_{\rm f}(r) = 
  \int \int \rho_{\rm P}(r_{\rm P}) \, \rho_{\rm T}(r_{\rm T}) \,
  v_{\rm eff}(E,\rho = \rho_{\rm P} + \rho_{\rm T}, 
	s = |\vec{r}+\vec{r_{\rm P}}-\vec{r_{\rm T}}|) \,
  d^3r_{\rm P} \, d^3r_{\rm T}
\label{eq:fold}
\end{equation}
where $\rho_{\rm P}$, $\rho_{\rm T}$ are the densities of projectile
and target, respectively, and $v_{\rm eff}$ is the effective
nucleon-nucleon interaction taken in the well-established DDM3Y
parametrization \cite{Sat79,Kobos84}. Details about the folding
procedure can be found in Refs.~\cite{Abele93,Atz96}. The folding
integral in Eq.~(\ref{eq:fold}) was calculated using the code DFOLD
\cite{DFOLD}.  

The resulting real part of the optical potential $V(r)$ is derived
from the folding potential $V_{\rm{f}}(r)$ by two minor modifications:
\begin{equation}
V(r) = \lambda \cdot V_{\rm f}(r/w)
\end{equation}
Firstly, the strength of the folding potential is adjusted by
the usual strength parameter $\lambda$ with $\lambda \approx 1.1 -
1.3$. This leads to volume integrals of the real potential [see
Eq.~(\ref{eq:vol})] of about
$J_R \approx 320 - 350$\,MeV\,fm$^3$ in the analyzed energy range
\cite{Mohr97,Atz96,Mohr2000a}. 
Secondly,
the densities of the $\alpha$ particle and the $^{92}$Mo nucleus were
derived from the experimentally known charge density distributions
\cite{devries87}, assuming identical proton and neutron distributions.
For $N \approx Z$ nuclei up to $^{90}$Zr ($Z = 40$, $N = 50$) this
assumption worked well \cite{Atz96}. However, to take the possibility
into account that the proton and neutron distributions are not
identical in the nucleus $^{92}$Mo ($Z = 42$, $N = 50$) a scaling
parameter $w$ for the width of the potential has been introduced,
which remains very close to unity. 

For a comparison of different potentials we use the
integral parameters volume integral per interacting
nucleon pair $J_R$ and the root-mean-square (rms) radius
$r_{\rm{rms,R}}$, which are given by:
\begin{eqnarray}
J_R & = &
	\frac{1}{A_{\rm P} A_{\rm T}} \, \int V(r) \, d^3r 
\label{eq:vol} \\
r_{\rm{rms,R}} & = &
	\left[ \frac{\int V(r) \, r^2 \, d^3r}
			{\int V(r) \, d^3r} \right] ^{1/2}
\label{eq:rms}
\end{eqnarray}
for the real part of the potential $V(r)$, and corresponding equations
hold for $W(r)$. $A_{\rm{P}}$ and $A_{\rm{T}}$ are the nucleon numbers
of projectile and target. Note that in the discussion of volume
integrals $J$ usually the negative sign is neglected; also in this
paper all $J$ values are actually negative.
The values for the folding potential $V_{\rm f}$ (with $\lambda = w = 1$)
are $J_R = 267.88~{\rm{MeV fm^3}}$ and $r_{\rm{rms,R}} = 4.989~{\rm{fm}}$.

The Coulomb potential is taken in the usual form of a 
homogeneously charged sphere where the Coulomb radius $R_{\rm C}$
is chosen identically with the rms radius of the
folding potential $V_{\rm f}$: $R_{\rm C} = r_{\rm{rms,R}} = 4.989~{\rm{fm}}$.

\subsection{The Imaginary Potential}
\label{subsec:imag}
Different parametrizations of the imaginary part of the optical
potential were chosen. Volume and surface Woods-Saxon (WS) potentials are
defined by the following equations
\begin{eqnarray}
W_{\rm{V}}(r) & = & W_0 \cdot f(x) \\
W_{\rm{S}}(r) & = & W_0 \cdot \frac{d}{dx}f(x)
\end{eqnarray}
with
\begin{equation}
f(x) = ( 1 + e^x )^{-1} {\mbox{~with~}} x = (r-R)/a
\end{equation}
The depth $W_0$, the radius parameter $R$, and the diffuseness $a$
have been adjusted to the experimental elastic scattering data.

Fourier-Bessel (FB) potentials are given by
\begin{equation}
W_{\rm{FB}}(r) = \sum_{k=1}^{n} 
	a_k \, \sin{(k \pi r/R_{\rm{FB}})} / (k \pi r/R_{\rm{FB}})
\end{equation}
with the cutoff radius $R_{\rm{FB}}$. Again, the Fourier-Bessel
coefficients $a_k$ are adjusted to the experimental data.

\subsection{Results and Continuous and Discrete Ambiguities}
\label{subsec:ambi}
\subsubsection{The 19\,MeV Data}
\label{subsubsec:19MeV}
The elastic scattering cross sections were calculated from optical
potentials with the computer code A0 \cite{A0}. The code allows a
variation of the potential parameters and determines the best-fit
values from a $\chi^2$ test.

\begin{figure}
\centerline{\epsfig{file=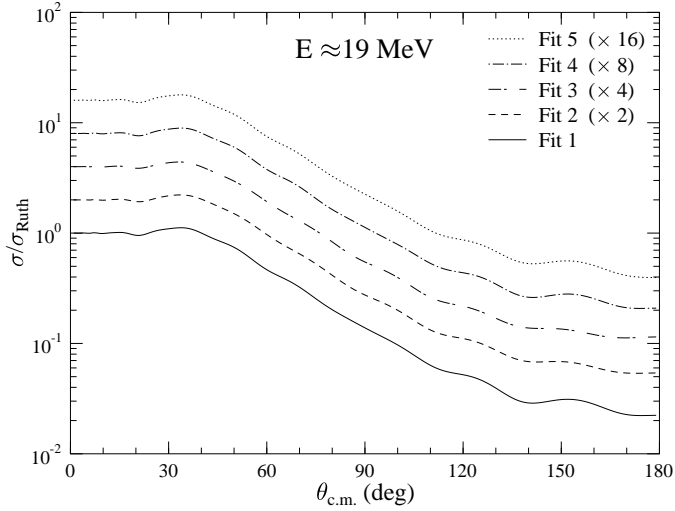,angle=0,width=9cm}}
\caption{
Calculated cross sections of $^{92}$Mo($\alpha$,$\alpha$)$^{92}$Mo
at $E_{\rm{c.m.}} \approx 19$\,MeV using five different
parametrizations of the imaginary part of the potential. These five
fits look very similar; 
however, fits 1, 4, and 5 have a significantly smaller $\chi^2$
(see Tables \ref{tab:final} and \ref{tab:fb}).
}
\label{fig:ambcont}
\end{figure}

\begin{figure}
\centerline{\epsfig{file=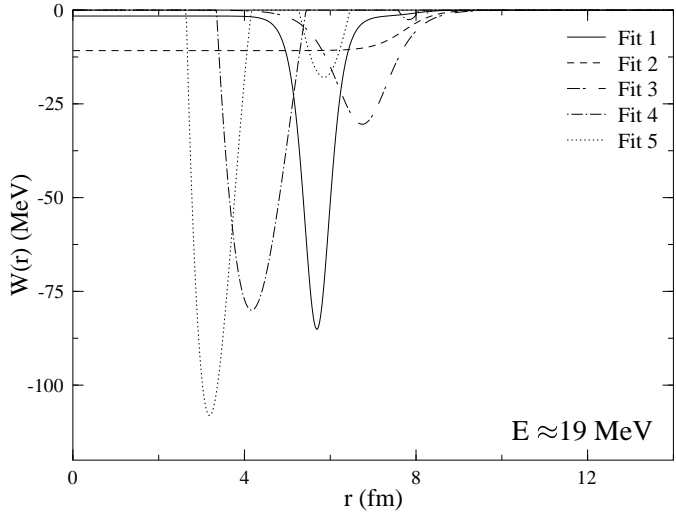,angle=0,width=9cm}}
\caption{
Imaginary potentials of fits 1-5 of $^{92}$Mo($\alpha$,$\alpha$)$^{92}$Mo
at $E_{\rm{c.m.}} \approx 19$\,MeV. The potential parameters are given
in Tables \ref{tab:final}, \ref{tab:ws} and \ref{tab:fb}. 
Further details see text.
}
\label{fig:imagpot}
\end{figure}

In the first analysis the potential parameters $\lambda$ and $w$ of
the real part were kept close to the expected values from the
systematic study \cite{Mohr97,Atz96,Mohr2000a}: $\lambda \approx 1.1 -
1.3$ and $w \approx 1.0$. Several parametrizations of the imaginary
potential were tested. It was found that different imaginary
potentials reproduced the experimental data with similar quality. Five
different fits are shown in Fig.~\ref{fig:ambcont} and the potential
parameters are given in Tables \ref{tab:final}, \ref{tab:ws} and \ref{tab:fb}.
It turns out that the real potential is practically identical in all
these fits, but the shape of the imaginary part shows strong
variations. The five imaginary potentials are shown in
Fig.~\ref{fig:imagpot}.

Since we want to determine the optical potential at astrophysically
relevant energies we have to extrapolate from the present
measurements. Because of the oscillating
behavior of the Fourier-Bessel potentials we decided to use the
combination of a volume and surface Woods-Saxon potential as basis for
the extrapolation. The $\chi^2$ obtained with this potential is
practically identical to the $\chi^2$ obtained from the Fourier-Bessel
potentials. The calculations with a pure volume Woods-Saxon or a pure
surface Woods-Saxon show significantly worse $\chi^2$ values.

\begin{figure}
\centerline{\epsfig{file=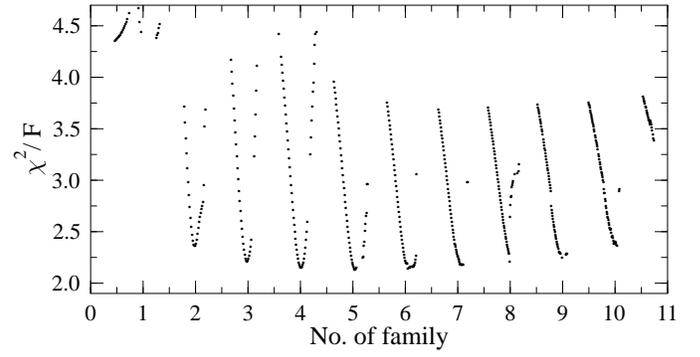,angle=0,width=9cm}}
\caption{
The variation of the strength parameter $\lambda$ of the real
potential in a wide range shows the so-called ``family
problem''. Several minima in $\chi^2$ can be found. Further detail see
text.
}
\label{fig:family}
\end{figure}

In a second analysis the strength parameter $\lambda$ and the width
$w$ of the real potential were varied in a wider range. It was found
that it is not possible to get a good fit to the data when the $w$
parameter deviates from 1.0 by more than a few per cent. However, a
variation of $\lambda$ leads to the known so-called ``family
problem''. It is possible to obtain comparable fits to the
experimental data with various $\lambda$ parameters. This phenomenon
was discussed in detail for a similar experiment in \cite{Mohr97}. 
In Fig.~\ref{fig:family} we
present the $\chi^2$ values which were obtained from the following
procedure: ($i$) the parameter $\lambda$ was varied from about 0.5 to
3.5; ($ii$) the width parameter $w$ and the imaginary part of the
potential (consisting of a combination of volume and surface
Woods-Saxon potentials) were adjusted to the experimental data for
each value of the strength parameter $\lambda$. One can clearly see
the families 2, 3, 4, and 5 as minima in $\chi^2$, corresponding to
$\lambda$ values of $0.81 - 1.52$. Note that the minima are more
pronounced than in the previous
$^{144}$Sm($\alpha$,$\alpha$)$^{144}$Sm experiment \cite{Mohr97}
because the ratio $E/V_C$ between the energy $E$ and the Coulomb
barrier $V_C$ is much higher in this $^{92}$Mo experiment.  

It is not possible to extract the correct family from these experimental
data only. But together with the systematic behavior of the volume
integrals found in \cite{Mohr97,Atz96,Mohr2000a} we can decide that
``family no.~4'' ($\lambda = 1.256$) should be used for the
description of the experimental scattering data and for the
extrapolation to astrophysically relevant energies (see
Sect.~\ref{sec:extra}). As mentioned above, family no.~4 
with $J_R \approx 340$\,MeV\,fm$^3$ corresponds to the values
of about $J_R \approx 320 - 350$\,MeV\,fm$^3$ which are expected from
the systematics of $\alpha$-nucleus potentials and also from other
systems \cite{Mohr2000c,Khoa00,Ogl00}. Neither family no.~3 with 
$J_R \approx 280$\,MeV\,fm$^3$ nor family no.~5 with 
$J_R \approx 400$\,MeV\,fm$^3$ fit into the systematics.
Numerical problems in the
fitting routine showed up for very shallow and very deep real
potentials, and a clear determination of families 1 and $6-10$ was not
possible.

\begin{figure}
\centerline{\epsfig{file=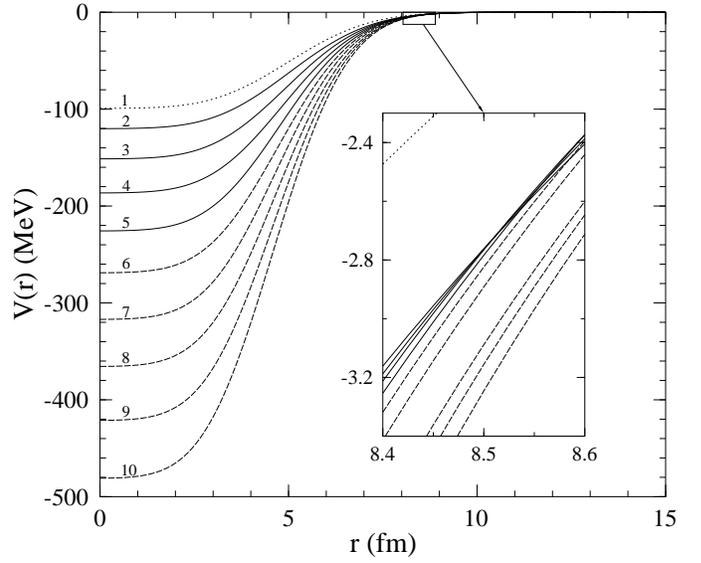,angle=0,width=9cm}}
\caption{ 
Real potentials for the different families 1-10 from
Fig.~\ref{fig:family}. The potentials from families 2-5 which are
well-defined as minima in $\chi^2$ (see Fig.~\ref{fig:family}) have
the same depth $V(r) = -2.66$\,MeV at the radius $r = 8.52$\,fm.  
}
\label{fig:onepoint}
\end{figure}

One further interesting fact has to be mentioned. The real potentials
corresponding to the families $1-10$ are shown in
Fig.~\ref{fig:onepoint}. The potentials from families $2-5$ which are
well-defined as minima in $\chi^2$ (see Fig.~\ref{fig:family}) have
the same depth $V(r) = -2.66$\,MeV at the radius $r =
8.52$\,fm. However, not all potentials which have this depth do
describe the data equally well; additionally, one has to find a
minimum in $\chi^2$ in Fig.~\ref{fig:family}. This behavior of a
so-called ``one-point potential'' has been observed in several
experiments, and the relevant radius has also been called ``sensitivity
radius'' (see e.g.~\cite{Sil01}); however, to our knowledge the
additional restriction of a significant minimum in $\chi^2$ has only
been observed in the analysis of the
$^{144}$Sm($\alpha$,$\alpha$)$^{144}$Sm data so far \cite{Mohr97}
which has been performed at a similar energy.

\subsubsection{The 13 and 16\,MeV Data}
\label{subsubsec:13and16MeV}
The procedure described in the previous Sect.~\ref{subsubsec:19MeV}
was repeated for the lower energies of 13 and 16\,MeV, and similar
results were obtained. The imaginary part of the potential
for all energies is taken as a combination of volume and surface
Woods-Saxon potentials. The potential parameters are listed in Table
\ref{tab:final}, and the calculations have been
compared to the experimental data already in Fig.~\ref{fig:results}.
However, the 16\,MeV data do not fit very well into the systematic
behavior shown in Fig.~\ref{fig:volume}. The potential extracted from the
16\,MeV data has larger uncertainties than at the other energies
because of experimental problems (see Sect.~\ref{subsec:angle}).

\subsubsection{30\,MeV Data from Literature}
\label{subsubsec:30MeV}
Two angular distributions at energies of about 30\,MeV are available
in literature \cite{Mats72,Mart68}. Unfortunately, both angular
distributions show systematic deviations between each other, and both
distributions have not been measured in the full angular range, but
in the ranges 
$15^\circ \le \vartheta_{\rm{c.m.}} \le 75^\circ$ \cite{Mats72} and
$15^\circ \le \vartheta_{\rm{c.m.}} \le 95^\circ$ \cite{Mart68}. If
one adjusts the potential parameters to these discrepant angular
distributions, discrepant optical potentials are obtained. 
The potential parameters are labeled in Table~\ref{tab:liter}. However,
the potentials extracted from the data of \cite{Mats72} fit into the
systematics, whereas the data of \cite{Mart68} do not fit. In both
cases the limited angular range restricts the sensitivity of the
potential parameters significantly. In Fig.~\ref{fig:volume} only the
volume integrals derived from \cite{Mats72} are shown.

\subsubsection{Backward Angle Excitation Function}
\label{subsubsec:exci}
The excitation function for $^{92}$Mo($\alpha$,$\alpha$)$^{92}$Mo has
been measured by Eisen {\it et al.} \cite{Eisen74} at $\vartheta_{\rm{lab}} =
170^\circ$ from 7 to 16\,MeV. We have calculated this excitation
function from our best-fit potentials at 13, 16, and 19\,MeV, and we
find excellent agreement between the experimental and the calculated
excitation function. 
The measured and the calculated excitation function at
$\vartheta_{\rm{lab}} = 170^\circ$ are shown in Fig.~\ref{fig:excit}.
The calculation with the 13\,MeV
potential underestimates the deviation from the
Rutherford cross section at higher energies. Such a behavior
can be expected because of the smaller volume integral of the imaginary
part in the 13\,MeV data corresponding to weaker absorption.
However, the measured scattering cross sections at one special
backward angle do not contain enough information to fix the optical
potential and its energy dependence.

\subsubsection{($\alpha$,n)-induced reactions}
\label{subsec:alphan}
A set of experimental data corresponding to the reaction 
$^{92}$Mo($\alpha$,$n$)$^{95}$Ru 
is available in \cite{Stro97} (Sect.IV B, Fig.6a). 
We have calculated the cross section from our model in the 
measured energy range, and found very good conformity between our 
calculations and the existing experimental data. 
However, the available data 
from the different experiments show discrepancies, 
which make it difficult to fix the energy dependence of the optical potential. 
Also the 
existing reaction data do not cover the astrophysically interesting energy
 range (between 7 and 9\,MeV, see Tab.~\ref{tab:gamow}), which would be 
helpful in order to confirm our predictions (see Sect.~\ref{sec:extra}).

\subsection{Discussion}
\label{subsec:disc}
\begin{figure}
\centerline{\epsfig{file=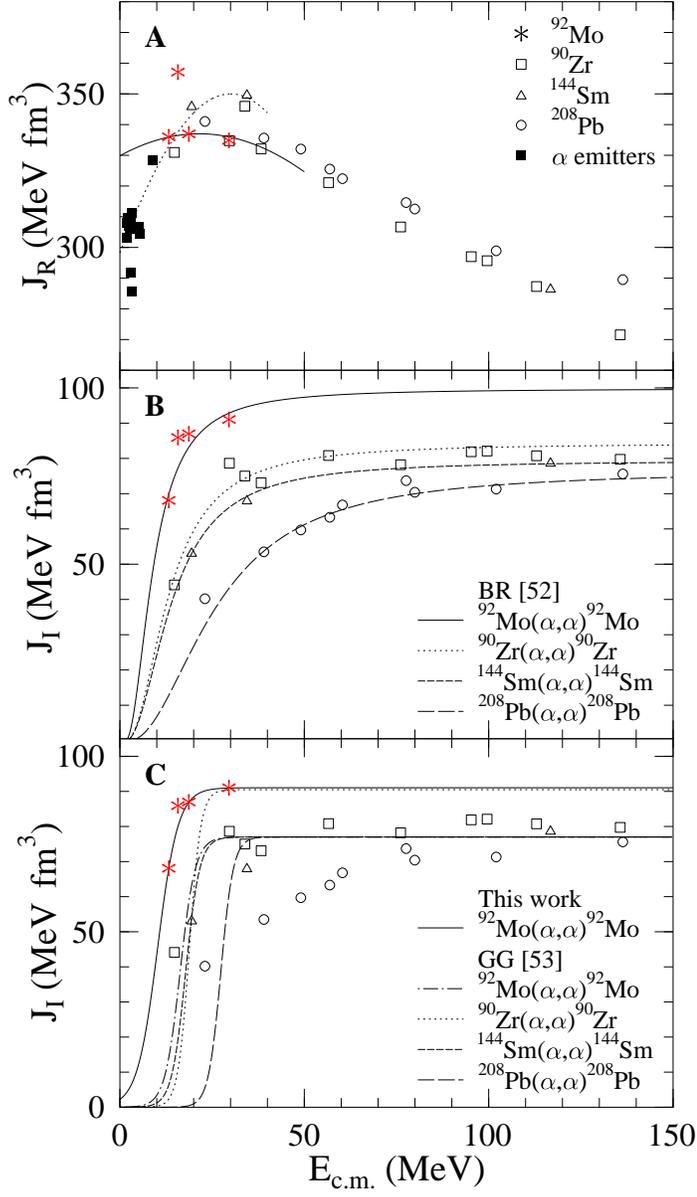,angle=0,width=9.21cm}}
\caption{ 
Volume integrals of the real (A, upper) and imaginary part
(B and C, center and lower diagram) of the optical potential derived from
$^{92}$Mo($\alpha$,$\alpha$)$^{92}$Mo scattering.  For comparison
volume integrals derived from $^{90}$Zr($\alpha$,$\alpha$)$^{90}$Zr,
$^{144}$Sm($\alpha$,$\alpha$)$^{144}$Sm, 
$^{208}$Pb($\alpha$,$\alpha$)$^{208}$Pb scattering 
\protect\cite{Atz96,Mohr97}, and from $\alpha$ emitters
\protect\cite{Mohr2000a} were added. The lines in the upper diagram
show Gaussian parametrizations of the new
$^{92}$Mo($\alpha$,$\alpha$)$^{92}$Mo data (full line) and from
Ref.~\protect\cite{Mohr2000a} (dotted line). 
The lines in the center diagram
show the results of BR parametrizations \protect\cite{Brown81} of the
imaginary part. In the lower diagram the lines are the result of a Fermi-like 
parametrization from 
Grama and Goriely (GG) 
\protect\cite{Gra98}, using the parameters proposed by the authors, and the 
ones derived in this work (full line). Details see text.
}
\label{fig:volume}
\end{figure}

\begin{figure}
\centerline{\epsfig{file=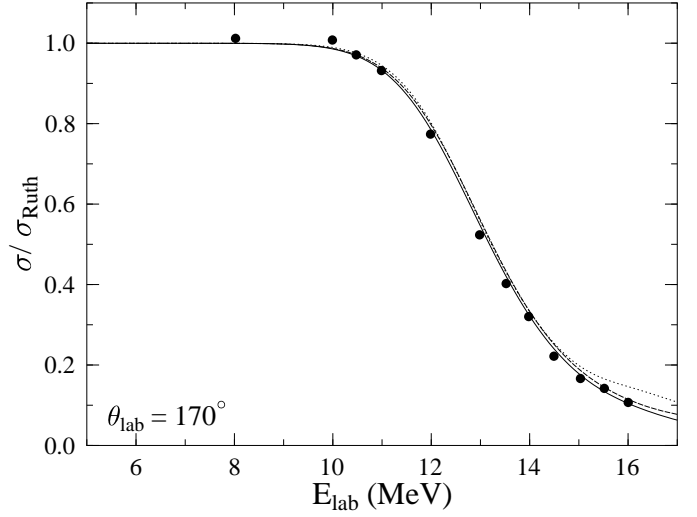,angle=0,width=9cm}}
\caption{ 
Excitation function of $^{92}$Mo($\alpha$,$\alpha$)$^{92}$Mo
scattering at $\vartheta_{\rm{lab}} = 170^\circ$ normalized to the
Rutherford cross section. The experimental
data are taken from Ref.~\protect\cite{Eisen74}. The dotted (dashed,
full) curves have been calculated from potentials which were adjusted
to the 13 (16, 19)\,MeV angular distributions.
}
\label{fig:excit}
\end{figure}

Volume integrals for various $\alpha$-nucleus potentials in a broad
range of masses and energies are shown in Fig.~\ref{fig:volume} for
the real (\ref{fig:volume}A) and the imaginary part
(\ref{fig:volume}B and \ref{fig:volume}C) of the optical potential.
The systematic behavior of volume integrals is also confirmed for
lighter target nuclei \cite{Abele93,Khoa01} and in
various other systems which have been analyzed recently
\cite{Mohr2000c,Khoa00,Ogl00}. For the extrapolation of the optical
potential to astrophysically relevant energies (Sect.~\ref{sec:extra})
parametrizations of the real and imaginary volume integrals are needed.

As can be seen from Fig.~\ref{fig:volume}A, there is only a weak energy
dependence of the real volume integral $J_R$ at energies below the
Coulomb barrier. 
As well as in Ref.~\cite{Mohr2000a}, a Gaussian parametrization is adjusted to 
the new data
\begin{equation}
J_R(E_{c.m.})\,=\,J_{R,0}\times \exp[-(E_{c.m.}-E_0)^2/\Delta^2]
\end{equation}
with $J_{R,0}$\,=\,337\,MeV\,fm$^3$, $E_0$\,=\,21.55\,MeV and 
$\Delta$\,=\,147.01\,MeV, leading to a curve (full line) which is somewhat 
flatter than the one proposed in Ref.~\cite{Mohr2000a} (dotted line).
The uncertainties for extrapolations to lower
energies are of the order of less than 5\,\% corresponding to about
$10-20$\,MeV\,fm$^3$.  Furthermore, the shape of the real potential is
given by the folding procedure (Sect.~\ref{subsec:fold}). This means
that the real part of the $\alpha$-nucleus optical potential can be
determined at energies below the Coulomb barrier with relatively small
uncertainties because ($i$) continuous ambiguities can be avoided
using the folding potential and ($ii$) discrete ambiguities can be
resolved from the systematic behavior of $\alpha$-nucleus potentials.

The situation for the imaginary part of the potential is much
worse. The volume integral $J_I$ of the imaginary part depends
strongly on the energy because many reaction channels open at energies
around the Coulomb barrier. Different parametrizations have been
proposed \cite{som98,Brown81,Gra98}. As an example we present the so-called
Brown-Rho (BR) parametrization \cite{Brown81}
\begin{equation}
J_I(E_{\rm{c.m.}}) =  \left\{ \begin{array}{rll}
	& \multicolumn{1}{c}{0} 
		& {\mbox{for~}} E_{\rm{c.m.}} \le E_0 \\
	& J_0 \cdot \frac{(E_{\rm{c.m.}} - E_0)^2}
		{(E_{\rm{c.m.}} - E_0)^2 + \Delta^2}
		& {\mbox{for~}} E_{\rm{c.m.}} > E_0 \\
	\end{array} \right.
\label{eq:br}
\end{equation}
with the excitation energy $E_0$ of the first excited
state. The saturation parameter $J_0$ and the
rise parameter $\Delta$ are adjusted to the
experimentally derived values. Another Fermi-like parametrization of the
imaginary volume integral, first introduced in Ref.\ \cite{som98}, reads
\begin{equation}
J_I(E_{\rm{c.m.}}) = \frac{J_0}{1+\exp{[(E^\ast-E_{\rm{c.m.}})/a^\ast]}}
\label{eq:gg}
\end{equation}
with a similar saturation value $J_0$ and the parameters $E^\ast$ and
$a^\ast$. The latter shape was also used for an attempt to derive a
global $\alpha$-potential \cite{Gra98}, with $E^\ast$ and $a^\ast$
depending on $E_0$. 
However, the line derived with the parameters given in \cite{Gra98} shows clear
 deviations from the new $^{92}$Mo data. Therefore, we have adjusted this 
Fermi-like function to the experimental data. 
Both parametrizations utilizing our fit parameters
are shown in 
Fig.~\ref{fig:volume}B (Brown-Rho) and ~\ref{fig:volume}C (Fermi-like) for 
our new $^{92}$Mo data. The parameters are
listed in Table \ref{tab:brgg}. In the following, we will always use the
parameters given in that table for the two descriptions
unless specified otherwise.

In general, the shapes of the BR and the Fermi parametrizations are quite
similar: there is a saturation value $J_0$ and a parameter that
describes the steep rise of $J_I$: $\Delta$ for BR and $a^\ast$ for
the Fermi shape. However, there are also important differences because the BR
parametrization leads to a somewhat flatter rise of $J_I$ than that of
the Fermi function, and
consequently, the extrapolation to lower energies is lower for the Fermi
parametrization than for BR. Consequences of these small differences
will be given in Sect.~\ref{subsec:rate}. Nevertheless, it should be
noted that the BR parametrization only contains two free parameters,
because $E_0$ is fixed, whereas the Fermi shape has three free
parameters, in principle.

The shown ambiguities do not allow to determine the shape of the
imaginary part. These ambiguities reduce the reliability of
extrapolations to lower energies. A more stringent determination of the
shape of the imaginary part requires extremely precise scattering data
over a wide range of energies. A scattering experiment at about
50\,MeV might help to reduce these uncertainties and to find the best
parametrization of the imaginary volume integrals.

\section{Extrapolation to Astrophysically Relevant Energies}
\label{sec:extra}

\subsection{The Astrophysically Relevant Energy for
($\gamma$,$\alpha$) reactions}
\label{subsec:E0}
The astrophysical decay rate $\tau^{-1}$ is given by
\begin{equation}
\tau^{-1}(T) =
  \int_0^\infty 
  c \,\, n_\gamma(E,T) \,\, \sigma_{(\gamma,\alpha)}(E) \,\, dE
\label{eq:tau}
\end{equation}
with the speed of light $c$, the cross section
$\sigma_{(\gamma,\alpha)}(E)$ of the ($\gamma$,$\alpha$) reaction, and
the photon density $n_\gamma(E,T)$ of a thermal photon bath at temperature $T$
\begin{equation}
n_\gamma(E,T) = 
  \left( \frac{1}{\pi} \right)^2 \,
  \left( \frac{1}{\hbar c} \right)^3 \,
  \frac{E^2}{\exp{(E/kT)} - 1}
\label{eq:planck}
\end{equation}
The integrand of Eq.~(\ref{eq:tau}) can be analyzed under the
assumption that the astrophysical S-factor of the reverse
($\alpha$,$\gamma$) reaction is constant: $S_{(\alpha,\gamma)}(E) =
const$. Then the maximum of the integrand in Eq.~(\ref{eq:tau}) is
found at the energy
\begin{equation}
E_0(\gamma,\alpha) = E_{\rm{thr}} + E_G^{1/3} \left(\frac{kT}{2}\right)^{2/3}
\end{equation}
with $E_G = 2\mu (\pi Z_P Z_T e^2/\hbar)^2$ and the threshold energy
$E_{\rm{thr}}$ for the ($\gamma$,$\alpha$) reaction. The most
effective energy $E_0(\gamma,\alpha)$ for ($\gamma$,$\alpha$)
reactions is given by the energy of the well-known Gamow window
$E_0(\alpha,\gamma)$ for the inverse ($\alpha$,$\gamma$) reaction plus the
separation energy $E_{\rm{thr}}$ of the $\alpha$ particle. Note that
the energy $E_0(\gamma,\alpha)$ is the energy of the photon, whereas
the energy $E_0(\alpha,\gamma)$ is the center-of-mass energy in the
system $^{92}$Mo - $\alpha$. The astrophysically relevant energies
for the system $^{92}$Mo - $\alpha$ are listed in Table
\ref{tab:gamow}.

In all astrophysical applications reaction rates are input only for
reactions with positive $Q$ value and the inverse rate is then computed
by applying detailed balance (see e.g.\ \cite{Rau2000}). That way,
numerical stability of the reaction network is guaranteed and the proper
equilibria of forward and reverse rates can be attained for a given
channel. The rates in the two directions depend linearly on each other
and thus the change of, say, the $\alpha$ potential equally influences
both, in our case the $\alpha$ capture as well as the
photodisintegration with $\alpha$ emission. This relation is valid
provided that stellar rates are used in both directions, accounting for
thermal excitation of the respective targets. Because of that fact, in the
following sections we make use of rate ratios so that the conclusions
apply to the forward and inverse rates as well.

\subsection{Extrapolation of the Optical Potential}
\label{subsec:extra}
As stated in Sect.~\ref{subsec:disc}, the extrapolation of the real
potential can be performed reliably leading to $J_R \approx
325$\,MeV\,fm$^3$ at astrophysically relevant energies with an
uncertainty of about 5\,\%. The corresponding strength parameter is
$\lambda \approx 1.2$. The width parameter was fixed at $w = 1.0$.

The extrapolation of the imaginary part was performed as follows. In a
first step the volume integral $J_I$ was determined from the BR and Fermi
parametrizations leading to $J_I = 23.9$\,MeV\,fm$^3$ (BR) and $J_I =
15.4$\,MeV\,fm$^3$ (Fermi) at $E_{\rm{c.m.}} = 5.81$\,MeV (corresponding
to $T_9 = 2.0$). The average of these values is $J_I = 19.6 \pm
4.2$\,MeV\,fm$^3$ which was used for the following calculations.

The shape of the potential was taken as sum of volume and surface
Woods-Saxon potentials where the radius parameter $R$ and the
diffuseness $a$ were estimated from the experimental data. The
contribution of the volume term to $J_I$ is assumed to be 30\,\%, and
the surface term contributes to 70\,\%. This ratio is determined from
the experimental scattering data at $E = 13$, 19, and 30\,MeV.
The effect of a variation of the relative contributions of volume and
surface term to $J_I$ will be discussed below. These and other
variations of the potentials allow an estimation
of the uncertainties of the calculated reaction rates.

\subsection{The $^{96}$Ru($\gamma$,$\alpha$)$^{92}$Mo reaction rate}
\label{subsec:rate}
The variation of the reaction rates when using various
potentials is shown in Tables \ref{tab:rateratio} and
\ref{tab:variation}. In Table \ref{tab:rateratio} the ratios of rates
obtained with the different potentials in respect to a standard rate
(taken from Refs.\ \cite{Rau2000,Rau2001} and using an $\alpha$ potential
from Ref.\ \cite{McF66}) are shown. As can be seen, the
rate calculated with the global potential of Ref.\ \cite{Gra98} is lower
by about two orders of magnitude than the standard rate. However, it was
already stated above that this potential does not describe the
$^{92}$Mo data at
higher energies. A simple equivalent square well potential
\cite{HWFZ} also yields a factor of 8 lower cross sections but neither does it
describe the data nor is it considered to be reliable for this application
\cite{woo2001}. When using the two potentials with the extrapolated
parameters for $T_9=2.0$ and $T_9=3.0$ from Table \ref{tab:extrapol}, a
reduction of the rate of about $40-50$\,\% is found. 

Case AB explores the dependence on the geometry of the potential. A
change in the geometry parameters of only $0.1-0.7$\,\% leads to a
variation in the ratio of $7-10$\,\% in the rate ratios which underlines
the importance of additional scattering experiments to determine the
shape of the imaginary optical potential. However, it should be mentioned
that case AB is not fully consistent within our approach because it
has a slightly different volume integral $J_I$ and rms radius
due to the unchanged depths of the volume and surface terms, but the
differences are only of the same order of magnitude as those in the geometry
parameters.

The sensitivity of the rates to variations in the extrapolated volume
integral $J_I$ and the relative contributions of volume and surface term
are studied in Table \ref{tab:variation}. Here the varied rates are
compared to the rate obtained in case A of Table \ref{tab:rateratio}.
The ratios are given in the temperature range $0.5\leq T_9 \leq 10.0$ in
order to show the temperature dependence of those effects although
strictly speaking the potential was derived assuming $T_9=2.0$.

The contribution of the surface term to $J_I$ was varied within a
reasonable range of $70\pm 20$\,\%. This resulted in a variation of
the rate of about $\pm 6$\,\%. Another uncertainty is introduced by the
fact that we assumed the extrapolated $J_I$ to be the mean between the
value obtained by the BR and Fermi parametrizations. Using the higher BR
value of $J_I=23.9$ MeV\,fm$^3$ increases the rate by 8\,\% while using
the lower value of $J_I=15.4$ MeV\,fm$^3$ a suppression by about 10\,\%
is obtained. Thus, the error introduced by the different shapes of the
parametrizations used for the extrapolation of $J_I$ to low energies is
dominating but still within satisfactory accuracy.

Closing this section we conclude that the recommended rates are case
A for $T_9=2.0$ and case B for $T_9=3.0$ from Table \ref{tab:rateratio} 
with an error of 16\,\%, mainly introduced by the ambiguities of the
extrapolation of the imaginary part down to the relevant energies.
The recommended rate is roughly a factor of two lower than the
standard rate given in previous tabulations \cite{Rau2000,Rau2001}.

\section{Summary}
\label{sec:summ}
We measured the elastic scattering cross section of
$^{92}$Mo($\alpha$,$\alpha$)$^{92}$Mo in a wide angular range
at energies of $E_{\rm{c.m.}} \approx$ 13, 16, and 19\,MeV.
Additionally, data from literature have been analyzed
\cite{Mats72,Mart68,Eisen74,Stro97}.
The real and imaginary parts of the optical potential for the
system $^{92}$Mo - $\alpha$ have been extracted from the data and
extrapolated down to the astrophysically relevant energies around and
below the Coulomb barrier. The result fits well into the known systematic
behavior of $\alpha$-nucleus folding potentials. The extrapolation of
the imaginary part is not unique but our study
shows that the use of two different energy dependencies introduces an
error in the obtained rate of
not more than 15\,\%. 

The derived stellar rates (for $^{92}$Mo($\alpha$,$\gamma$)$^{96}$Ru as
well as $^{96}$Ru($\gamma$,$\alpha$)$^{92}$Mo) are $50-60$\,\% of the
rates given in Refs.\ \cite{Rau2000,Rau2001} at stellar temperatures
$T_9=2.0-3.0$. Assuming the $^{96}$Ru($\gamma$,n)$^{95}$Ru rate to remain
unchanged, this would lead to a corresponding decrease in the abundance ratio 
$Y_{^{92}{\rm Mo}}/Y_{^{94}{\rm Mo}}$ with respect to an abundance ratio 
calculated with the previous rate in the $\gamma$-process
(as e.g.\ in \cite{heg00,rau_apj01}). It is interesting to note that
many network calculations for the $\gamma$-process 
\cite{Woo78,Ray90,Ray95,Pra90,How91}
show an overproduction of $^{92}$Mo relative to $^{94}$Mo
which may be reduced by the results of this work. However,
as mentioned in the introduction,
a complete analysis has not only to follow the $\gamma$-process
consistently but also to account for the possible $rp$- and $s$-process
contributions. This is beyond the scope of this paper.

\acknowledgments
We would like to thank the cyclotron team of ATOMKI for the
excellent beam during the experiment. Two of us (P.~M., M.~B.)
gratefully acknowledge the kind hospitality at ATOMKI. 
Zs.~F.~is a Bolyai fellow. T.~R.~is
a PROFIL professor (Swiss NSF grant 2124-055832.98). This work was
supported by OTKA (T034259),
the Swiss NSF (grant 2000-061822.00), and the
U.S. NSF (contract NSF-AST-97-31569).

\newpage

%\newpage

%\newpage

\end{multicols}
\newpage

\begin{table}
\caption{
	Potential parameters of the imaginary part of
	the optical potential (combination of volume and surface WS 
	parametrizations), 
	derived from the angular distribution of
	$^{92}$Mo($\alpha$,$\alpha$)$^{92}$Mo
	at $E = $13, 16 and 19\,MeV,
	and its
	integral potential parameters $J$ and $r_{\rm{rms}}$
	of their real and imaginary parts. 
}
{
\begin{tabular}{cccccccccc}
fit &
$E$     & $W_V~({\rm{MeV}})$    
	& $R_V~({\rm{fm}})$     
	& $a_V~({\rm{fm}})$ &   
	& $W_O~({\rm{MeV}})$ 
	& $R_O~({\rm{fm}})$
	& $a_O~({\rm{fm}})$
	& \\
\hline
%
% fit 4b
1 & 19  & -1.584 & 1.7667 & 0.2659 & & 334.25 & 1.2605 & 0.2073 & \\
- & 16  & -9.558 & 1.668 & 0.248 & & 308.20 & 1.348 & 0.099 \\
- & 13  & -5.128 & 1.656 & 0.002 & & 467.06 & 1.369 & 0.071 \\
\hline\hline
\\
\hline\hline
fit &
$E$     & $\lambda$     & $w$   
	& $J_R$ & $r_{\rm{rms,R}}$
	& $J_I$ & $r_{\rm{rms,I}}$
	& $\chi^2/F$ \\
 	& (MeV)   &               &       
	& $({\rm{MeV fm^3}})$   & $({\rm{fm}})$
	& $({\rm{MeV fm^3}})$   & $({\rm{fm}})$
	& \\
\hline
%
% fit 4b
1 & 19 & 1.257 & 1.003  & 337.2 & 4.991 & 86.2 & 5.806 & 2.15 \\
- & 16 & 1.346 & 0.9974 & 357.8 & 4.976 & 85.9 & 5.992 & 4.84 \\
- & 13 & 1.352 & 0.9758 & 336.7 & 4.869 & 67.3 & 6.043 & 1.26 \\
\end{tabular}
}
\label{tab:final}
\end{table}

\begin{table}
\caption{
	Potential parameters of the imaginary part 
	of the optical potential (volume and surface WS 
	parametrization), 
	derived from the angular distribution of
	$^{92}$Mo($\alpha$,$\alpha$)$^{92}$Mo
	at $E = $19\,MeV, and its
	integral potential parameters $J$ and $r_{\rm{rms}}$
	of their real and imaginary parts.	
}
{
\begin{tabular}{ccccccccc}
fit     & $W_V~({\rm{MeV}})$    
	& $R_V~({\rm{fm}})$     
	& $a_V~({\rm{fm}})$       
	&    
	& $W_O~({\rm{MeV}})$ 
	& $R_O~({\rm{fm}})$
	& $a_O~({\rm{fm}})$
	& \\
\hline
2       & -10.806 & 1.7116 & 0.3601 & & & & & \\
3	& & & & & 121.76 & 1.4947 & 0.4206 & \\
\hline\hline
\\
\hline\hline
fit     & $\lambda$     & $w$   
	& $J_R$ & $r_{\rm{rms,R}}$
	& $J_I$ & $r_{\rm{rms,I}}$
	& $\chi^2/F$ \\
No.     &               &       
	& $({\rm{MeV fm^3}})$   & $({\rm{fm}})$
	& $({\rm{MeV fm^3}})$   & $({\rm{fm}})$
	& \\
\hline
%
% fit 6b
2       & 1.237               & 1.010
	& 341.1               & 5.037        & 57.9 & 6.131    & 3.67 \\
% fit 9b
3       & 1.188               & 1.021
	& 338.9               & 5.095        & 80.6 & 6.957    & 4.28 \\
\end{tabular}
}
\label{tab:ws}
\end{table}

\begin{table}
\caption{
	Potential parameters of the imaginary part of
	the optical potential (FB para- metrization), 
	derived from the angular distribution of
	$^{92}$Mo($\alpha$,$\alpha$)$^{92}$Mo
	at $E = $19\,MeV,
	and its
	integral potential parameters $J$ and $r_{\rm{rms}}$
	of their real and imaginary parts.
}
{
\begin{tabular}{cccccccccc}
fit     & $R_{\rm{FB}}~({\rm{fm}})$
	& $a_1$ & $a_2$ & $a_3$ & $a_4$
	& $a_5$ & $a_6$ & $a_7$ \\	
\hline
%
% fit 9b
4       & 12.8 & 154.94 & -311.98 & 839.41 & -325.97 & 880.41 & - & - \\
% fit 10b
5       & 12.0 & 167.94 & -329.11 & 1103.43 & -592.10 & 1666.61 & -346.79 &
 1118.06 \\
\hline\hline
\\
\hline\hline
fit     & $\lambda$     & $w$ &  
	& $J_R$ & $r_{\rm{rms,R}}$
	& $J_I$ & $r_{\rm{rms,I}}$
	& $\chi^2/F$  \\
No.     &               & &       
	& $({\rm{MeV fm^3}})$   & $({\rm{fm}})$
	& $({\rm{MeV fm^3}})$   & $({\rm{fm}})$
	& \\
\hline
4       & 1.272               & 0.998 &
	& 338.8               & 4.979        & 68.2 & 4.524   & 2.23 \\
5      & 1.287    		& 0.991 &
       & 336.0	&  4.947	& 53.9  & 4.319 & 2.14 \\
\end{tabular}
}
\label{tab:fb}
\end{table}

\begin{table}
\caption{
	Potential parameters of the imaginary part of
	the optical potential (combination of volume and surface WS 
	parametrizations), 
	derived from the angular distribution of
	$^{92}$Mo($\alpha$,$\alpha$)$^{92}$Mo
	from Refs.~\protect\cite{Mats72} and \protect\cite{Mart68},
	and its
	integral potential parameters $J$ and $r_{\rm{rms}}$
	of their real and imaginary parts.
}
{
\begin{tabular}{ccccccc}
Ref.    & $W_V~({\rm{MeV}})$    
	& $R_V~({\rm{fm}})$     
	& $a_V~({\rm{fm}})$       
	& $W_O~({\rm{MeV}})$ 
	& $R_O~({\rm{fm}})$
	& $a_O~({\rm{fm}})$\\
\hline
\cite{Mats72}& -4.91 & 1.78 & 0.40 & 183.13 & 1.13 & 0.36 \\
\cite{Mart68}& -3.51 & 1.72 & 0.95 & 337.06 & 1.26 & 0.23 \\
\hline\hline
\\
\hline\hline
Ref.    & $\lambda$     & $w$   
	& $J_R$ & $r_{\rm{rms,R}}$
	& $J_I$ & $r_{\rm{rms,I}}$\\
No.     &               &        
	& $({\rm{MeV fm^3}})$   & $({\rm{fm}})$
	& $({\rm{MeV fm^3}})$   & $({\rm{fm}})$\\	
\hline
\cite{Mats72}& 1.19 & 1.022 & 334.76 & 5.098 & 90.95 & 5.704 \\
\cite{Mart68}& 1.15 & 1.014 & 315.49 & 5.056 & 107.91 & 6.014 \\
\end{tabular}
}
\label{tab:liter}
\end{table}

\begin{table}
\caption{
	Extrapolated values of the potential paramters 
	of the imaginary part of the optical potential at
	the astrophysically relevant energies $E_0 = 5.8$\,MeV ($T_9 =
	2.0$) and $E_0 = 7.6$\,MeV ($T_9 = 3.0$),
	and its integral potential parameters $J$ and $r_{\rm{rms}}$
	of their real and imaginary parts.
}
{
\begin{tabular}{ccccccccc}
$E$     & $W_V~({\rm{MeV}})$    
	& $R_V~({\rm{fm}})$     
	& $a_V~({\rm{fm}})$ &   
	& $W_O~({\rm{MeV}})$ 
	& $R_O~({\rm{fm}})$
	& $a_O~({\rm{fm}})$
	& \\
\hline
7.6 & -1.466 & 1.717 & 0.268 & & 36.86 & 1.295 & 0.419 \\
5.8 & -1.091 & 1.720 & 0.270 & & 27.44 & 1.297 & 0.420 \\
\hline\hline
\\
\hline\hline
$E$     & $\lambda$     & $w$   
	& $J_R$ & $r_{\rm{rms,R}}$
	& $J_I$ & $r_{\rm{rms,I}}$
	& $\chi^2/F$ \\
(MeV)   &               &       
	& $({\rm{MeV fm^3}})$   & $({\rm{fm}})$
	& $({\rm{MeV fm^3}})$   & $({\rm{fm}})$
	& \\
\hline
7.6 & 1.219 & 1.000 & 327.1 & 4.991 & 26.2 & 6.085 & - \\
5.8 & 1.209 & 1.000 & 324.3 & 4.991 & 19.6 & 6.095 & - \\
\end{tabular}
}
\label{tab:extrapol}
\end{table}

\begin{table}
\caption{
Parameters of the BR and Fermi parametrizations of the imaginary volume
integral $J_I$ for
$^{92}$Mo($\alpha$,$\alpha$)$^{92}$Mo.
}
{
\begin{tabular}{cccc}
parametrization & saturation value	
	& rise parameter    
	& other parameters \\
\hline
BR	& $J_0=99.8$\,MeV\,fm$^3$	
	& $\Delta = 7.68$\,MeV	
	& $E_0 = 1.51$\,MeV	\\
Fermi	& $J_0=91.0$\,MeV\,fm$^3$	
	& $a^\ast = 2.78$\,MeV	
	& $E^\ast = 10.24$\,MeV	\\
\end{tabular}
}
\label{tab:brgg}
\end{table}

\begin{table}
\caption{
Most effective energies $E_0$ for the
$^{92}$Mo($\alpha$,$\gamma$)$^{96}$Ru and the
$^{96}$Ru($\gamma$,$\alpha$)$^{92}$Mo reactions.
}
{
\begin{tabular}{ccc}
$T_9$	& $E_0(\alpha,\gamma)$    & $E_0(\gamma,\alpha)$ \\
\hline
2.0	& 5.81	& 7.51	\\
2.5	& 6.75	& 8.44	\\
3.0	& 7.62	& 9.31	\\
\end{tabular}
}
\label{tab:gamow}
\end{table}

\begin{table}
\caption{Ratio $\xi=r_x/r_0$ of the astrophysical reaction rates $r_x$ 
obtained with different imaginary potentials to a standard rate $r_0$ 
\protect\cite{Rau2000,Rau2001}.}
\begin{tabular}{rrrrrr}
\multicolumn{1}{c}{$T_9$}&\multicolumn{1}{c}{GG\tablenotemark[1]}&
\multicolumn{1}{c}{ESW\tablenotemark[2]}&
\multicolumn{1}{c}{A \tablenotemark[3] ($T_9=2$)}&
\multicolumn{1}{c}{B \tablenotemark[4] ($T_9=3$)}&
\multicolumn{1}{c}{AB \tablenotemark[5]}\\
\hline
  2.0 & 0.014 & 0.121 & 0.453 & 0.497 & 0.503 \\
  3.0 & 0.012 & 0.140 & 0.546 & 0.579 & 0.585
\end{tabular}
\tablenotetext[1]{Potential from Ref.\ \protect\cite{Gra98}}
\tablenotetext[2]{Equivalent square well potential, e.g.\ as in Ref.\
\protect\cite{HWFZ}}
\tablenotetext[3]{Imaginary part from the extrapolated values at $E=5.8$
MeV in Tab.\ \protect\ref{tab:extrapol}.}
\tablenotetext[4]{Imaginary part from the extrapolated values at $E=7.6$ 
MeV in Tab.\ \protect\ref{tab:extrapol}.}
\tablenotetext[5]{Potential depths are from A, the geometry parameters from B.}
\label{tab:rateratio}
\end{table}

\begin{table}
\caption{Variation of the imaginary part of the potential derived for
$T_9=2.0$. Rate ratios are shown in respect to the rate obtained with
the parameters for $E=5.8$ MeV ($J_I=19.6$ MeV\,fm$^3$, 70\% surface
contribution) from Tab.\ \protect\ref{tab:extrapol}.}
\begin{tabular}{ccccrrrr}
&\multicolumn{2}{c}{Variation of $J_I$}&&
\multicolumn{4}{c}{Surface Contribution}\\
\cline{2-3} \cline{5-8}
T&\multicolumn{1}{c}{$J_I=$}&\multicolumn{1}{c}{$J_I=$}
&&\multicolumn{4}{c}{$J_I = 19.6$ MeV\,fm$^3$} \\
10$^9$ K& \multicolumn{1}{c}{15.4 MeV\,fm$^3$} &\multicolumn{1}{c}{23.9
MeV\,fm$^3$}& &\multicolumn{1}{c}{90\%}&\multicolumn{1}{c}{80\%}&
\multicolumn{1}{c}{60\%}&\multicolumn{1}{c}{50\%}\\
\hline
 0.5 & 0.942 & 1.077 & &0.952 & 0.976 & 1.019 & 1.043\\
 1.0 & 0.949 & 1.070 & &0.949 & 0.975 & 1.032 & 1.057\\
 1.5 & 0.913 & 1.078 & &0.937 & 0.968 & 1.029 & 1.057\\
\hline
 2.0 & 0.902 & 1.077 & &0.937 & 0.969 & 1.029 & 1.056\\
\hline
 2.5 & 0.918 & 1.062 & &0.945 & 0.974 & 1.026 & 1.049\\
 3.0 & 0.945 & 1.050 & &0.958 & 0.981 & 1.020 & 1.040\\
 3.5 & 0.968 & 1.027 & &0.966 & 0.984 & 1.016 & 1.030\\
 4.0 & 0.991 & 1.013 & &0.977 & 0.988 & 1.010 & 1.021\\
 4.5 & 1.007 & 1.001 & &0.983 & 0.992 & 1.006 & 1.014\\
 5.0 & 1.019 & 0.995 & &0.989 & 0.995 & 1.005 & 1.009\\
 6.0 & 1.028 & 0.991 & &1.000 & 1.000 & 1.000 & 1.009\\
 7.0 & 1.031 & 0.985 & &0.999 & 0.999 & 1.000 & 1.000\\
 8.0 & 1.026 & 0.987 & &1.000 & 1.000 & 1.000 & 0.996\\
 9.0 & 1.020 & 0.990 & &1.004 & 1.002 & 0.998 & 0.996\\
10.0 & 1.011 & 0.994 & &1.005 & 1.002 & 0.998 & 0.994
\end{tabular}
\label{tab:variation}
\end{table}

%\end{multicols}

\end{document}